# Features of stimulated luminescence of solid nitrogen

M.A. Bludov[1], I.V. Khyzhniy[1], S.A. Uyutnov[1], G.B. Gumenchuk[2] E.V. Savchenko[1]

[1]*B. Verkin Institute for Low Temperature Physics and Engineering of NAS of Ukraine, 47 Nauky Ave., Kharkiv, 61103, Ukraine*
[2]*Lehrstuhl für Physikalische Chemie TUM, Garching b. München 85747, Germany*

e-mail: savchenko@ilt.kharkov.ua

**Abstract**

Recent results on the study of spontaneous and stimulated luminescence of solid nitrogen in the near-infrared NIR range are presented. Irradiation was performed with an electron beam of subthreshold energy in the dc mode. Three series of experiments were performed: (i) measurement of cathodoluminescence CL at different electron energies on samples of different thicknesses, (ii) measurements of thermally stimulated luminescence TSL in combination with thermally stimulated exoelectron emission TSEE from pre-irradiated samples and (iii) recording of non-stationary luminescence curves NsL at selected wavelengths during gradual heating of samples under an electron beam. Three emission bands were recorded in the NIR TSL spectra of solid $N_2$: 794, 802, and 810 nm which form the γ-group. The band at 810 nm in stimulated luminescence was detected for the first time. The positions of all three spectral features coincide in the spectra of spontaneous and stimulated luminescence, as evidenced by a comparison of the CL spectrum recorded at 5 K with the TSL spectrum recorded at the TSL maximum at 16 K. A comparison of the CL spectra obtained under different conditions showed that there is no complete correlation in the behavior of emission from the $^2D$ state of the N atom and the γ-group, which could be expected in the case of the formation of the γ-band emitting centers via electron attachment to the $N(^2D)$ atom as it was suggested in [29]. The glow curves measured for these 3 bands were found to correlate with each other and with the TSEE yield. This finding indicates common origin of these bands and their connection with the neutralization reaction. The correlation of the γ–band NsL in the range of low temperatures (5-20 K) with the NsL measured at the 0-4 band of the $a'^1\Sigma_u^- \rightarrow X^1\Sigma_g^+$ transition, which is the "fingerprint" of the tetranitrogen cation $N_4^+$ [35], points to possible connection of the γ–band with the neutralization of $N_4^+$.

**Introduction**

The ongoing interest in research of radiation effects in solid Nitrogen and $N_2$–containing ices is driven by its significant role in astrophysics [1-12], its prospects for generation of polynitrogen compounds considered as environment-friendly high energy-density materials (HEDM) [13-17] and importance of nitrogen in radiation physics and chemistry. Active research of the problem of charged species dynamics and their role in a variety of radiation-induced phenomena has been undertaken in recent years and its trends were reviewed in [18]. One of the related questions concerns identification of, the so-called γ-line, situated in the near infrared (NIR) range. This line was observed during the irradiation of solid nitrogen with electrons [19-22] and fast atoms [23] and was detected during the condensation of samples from the discharge and their subsequent heating [24-29]. Despite the numerous experiments an assignment of this line remained under question for a long time. The most detailed study of the γ-line was performed using the technique of growing impurity helium condensates (IHC) by injecting a helium gas jet with impurities ($N_2$ and a mixture of $N_2$ with Ne or Ar or Kr), which had previously passed through a discharge, into superfluid helium (HeII) [29]. The γ-line position depended on the content of nanoclusters and for $N_2$, Ar and Ne was varied around 793 nm according [29]. Relation of this line to interaction of nitrogen species with electrons was established in [22, 29, 30]. It is necessary to emphasize that there are two fundamentally different channels of interaction of nitrogen centers with electrons, viz. (i) the attachment of electrons to neutral centers with anion formation and (ii) the recombination of electrons with positively charged centers, which results in their neutralization. The channel (i) was considered by the authors of [29]. It is known that nitrogen atom in the ground state has negative electron affinity $E_a$=-0.07 eV [31]. However, for the excited $^2D$ term a non-relativistic electron affinity is predicted to be of 0.871(2) eV [32, 33]. Taking into account that the position of $^2D$ state is 2.3838 eV above that of the ground-state $^4S_{3/2}$, the energy of the excited $^1D$ term of $N^−$ was estimated to be 1.513 eV above the ground-state energy of nitrogen atom and about 1.44 eV above the ground state $^3P$ of nitrogen anion. Based on these theoretical calculations the authors of [29] suggested that the γ–line appears as a result of the electron attachment to the metastable $N(^2D)$ atom forming nitrogen anion $N^−$ in the excited state $^1D$: $e^- + N(^2D) \rightarrow N^−(^1D)$, and assigned the γ–line to bound-bound transitions of $N^-$ center: $N^−(^1D) \rightarrow N^−(^3P) + h\nu$ (the γ-line). However, this hypothesis does not explain existence of the red satellite of the γ–line near 802 nm which was observed in [29] and in [19, 30] under excitation by an electron beam. Moreover, finding the second satellite of the γ–line at 810 nm in the spectrum of spontaneous luminescence [30] calls into question the identification of the γ–line as the emission of nitrogen anion $N^−$. Note that the γ–line and two satellites in the CL spectrum which are characterized by similar behavior form molecular series with spacing between adjacent vibrational energy levels of the ground state of 125 cm$^{-1}$ and 123 cm$^{-1}$. Real-time correlated measurements of spectrally resolved thermally stimulated luminescence TSL and exoelectron emission performed in the temperature range of the nitrogen α-phase existence and in the Ne matrix in the temperature range of 5-15 K established correlation of the γ-line with TSEE currents and recombination emissions of $O^+$, $N^+$, and $N_2^+$ ions:

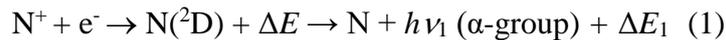
$N^+ + e^- \rightarrow N(^2D) + \Delta E \rightarrow N + h\nu_1$ (α-group) $+ \Delta E_1$ (1)

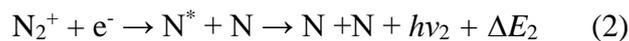
$N_2^+ + e^- \rightarrow N^* + N \rightarrow N + N + h\nu_2 + \Delta E_2$ (2)

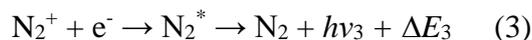
$N_2^+ + e^- \rightarrow N_2^* \rightarrow N_2 + h\nu_3 + \Delta E_3$ (3)

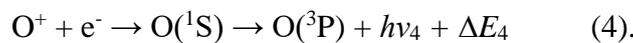
$O^+ + e^- \rightarrow O(^1S) \rightarrow O(^3P) + h\nu_4 + \Delta E_4$ (4).

These findings point to the connection of the γ-line with another channel of interaction of nitrogen centers with electrons, viz. (ii) – neutralization reactions. It is noteworthy that even for the O atom, characterized by a high positive electron affinity $E_a$=1.46 eV [34], the neutralization channel of relaxation is predominant. One more neutralization channel was revealed in solid nitrogen [35] from the spectrally resolved TSL and NsL study in the VUV range in combination with TSEE measurements. „Fingerprints" of $N_4^+$ were found as products of the dissociative recombination reaction stimulated by electron detrapping upon heating:

$$N_4^+ + e^- \rightarrow N_4^{**} \rightarrow N_2^*(a'^1\Sigma_u^-) + N_2^*(a'^1\Sigma_u^-) + \Delta E \rightarrow N_2 + N_2 + 2h\nu_5 + \Delta E_5 \quad (5)$$

By this relaxation scenario, so-called „cage exit", the cation $N_4^+$ recombines with electron followed by dissociation of the transient products $N_4^{**}$ into $N_2^*$ which relax radiatively into the ground state. This scenario is realized if there is no barrier to dissociation or this barrier is too low.

If there is a barrier, the excited $N_2^*$ molecule after dissociation interacts with a ground state $N_2$ molecule from the environment to form $N_4^*$ and then $N_4$ in the ground state via radiative transition:

$$N_4^+ + e^- \rightarrow N_4^{**} \rightarrow N_4^* \rightarrow N_2^* + N_2 + \Delta E \rightarrow N_4^* \rightarrow N_4 + h\nu_4 + \Delta E_6 \quad (6)$$

This scenario, called „cage effect", is observed in a number of radiation-induced reactions [36, 37]. The question remains whether this scenario is in operation in solid nitrogen. As it was noted, connection of the γ-line with neutralization reaction was established in [30] and it was suggested that the reaction corresponds to the case (5) because the experimental energy of radiative transition (1.56 eV) appeared to be very close to the predicted (1.55 eV) for the transition from the lowest excited state $1B_{3u}$ of the neutral tetranitrogen compound $N_4$ of $D_{2h}$ configuration [38]. However, the experimental harmonic frequency appeared to be much lower than the lowest calculated one, which does not allow this band identification as the radiative $1\,^1B_{3u} \rightarrow\,^1A_g$ transition of the neutral compound $N_4$ of $D_{2h}$ configuration to be considered definitive. In view of that the assignment of the γ-line and its satellites remained the subject of further experimental and theoretical studies.

Motivated by this open question, we performed the additional experiments on the study of spontaneous and stimulated luminescence of nitrogen solids described in this paper. Irradiation was performed in dc regime with an electron beam of subthreshold energy. First of all, we checked if there is correlation in behavior of the α-group related to the $^2D \rightarrow\,^4S$ transition of nitrogen atom and the γ-group. CL spectra of different thickness samples were recorded in the visible and NIR range, as well as at different excitation energy. Comparison of CL spectra obtained under different conditions showed the absence of a complete correlation in the behavior of radiation from the $^2D$ state of the N atom and the γ-line, which could be expected in the case of the formation of γ-line emitting centers due to the attachment of an electron to the $N(^2D)$ atom. Increasing the irradiation time made it possible to register a new NIR band at 810 nm (the second satellite of the γ-line) in the spectra of TSL. The positions of all three spectral features coincide in the spectra of spontaneous and stimulated luminescence. Measurement of the TSL curves at 802 and 810 nm showed a correlation with the TSL curve measured on the γ-line at 794 nm and the yield of the TSEE. The similar behavior of all these lines in the stimulated luminescence (TSL) and their correlation with the stimulated current (TSEE) indicate common origin of these lines and their connection with the neutralization reaction. An additional argument in favor of the connection of the γ-line with $N_4^+$ neutralization was provided by the measurements of spectrally resolved NsL curves in the VUV and NIR ranges. It was established that the γ line NsL in the low temperature range (5-20 K) correlates with the NsL measured in the 0-4 band of the $a'^1\Sigma_u^- \rightarrow X^1\Sigma_g^+$ transition, which is a "fingerprint" of the neutralization of the tetranitrogen cation $N_4^+$ [35].

**Experimental**

The experimental approach to study radiation effects in ices and has been described in detail in sect. 7 of ref. [39] and, in particular, in solid nitrogen in [18]. Therefore, we will briefly describe the course of the experiments. High-purity $N_2$ (99.995%) was used to grow films of different thicknesses (from 30 nm to 300 μm) by deposition from the gas phase on a cold Cu substrate mounted on a liquid He cryostat finger. Samples were grown and irradiated at 5 K. The sample heating under electron beam did not exceed 0.7 K. The $N_2$ solid films temperature was controlled with a Si sensor. The open surface enabled us to perform spectral measurements in a wide range (from VUV up to NIR) and detect stimulated electron emission. The irradiation was carried out in dc mode with an electron beam of subthreshold energy to exclude the knock-on defect formation and sputtering. Taking into account that the threshold electron beam energy $E_{thr}$ for production of defects in solid $N_2$ via the knock-on mechanism is about 1.5 keV we used beams with energies in the range 500-1500 eV. The beam covered the icy film with an area of 1 cm$^2$. The CL spectra from 100 nm up to 900 nm were detected concurrently with two spectrometers. Note, that spectra were not corrected for the spectral sensitivity of the optical system.

On completing the exposure we detected an "afteremission" current. This phenomenon was observed in [40] which reported first detection of TSEE from solid nitrogen pre-irradiated with an electron beam. Measurements of thermally stimulated relaxation emissions TSL and TSEE from solid nitrogen were started when the "afteremission" current and afterglow had decayed to essentially zero. In most experiments we used heating with a constant rate of 5 K min$^{-1}$. Stimulated currents were detected with an electrode kept at a small positive potential $V_F$=+9 V and connected to the current amplifier. Due to high mobility of electrons in solid nitrogen in the range of α-phase existence [41] TSEE measurements provide information not only on surface-related processes but also on the processes occurring in deeper layers of the films. Given the sensitivity of TSEE, TSL, and other phenomena to sample structure, impurity concentration, and other variables, these measurements were performed in a correlated manner on the same sample. Additional series of experiments was carried out using the method of nonstationary luminescence (NsL) developed by our group [42]. The idea behind of this "pump-probe" technique is the dosed release of electrons from traps, driving by heating under beam, to neutralize positively charged species contributing accordingly to NsL. The combination of optical and current activation spectroscopy methods made it possible to track various relaxation channels and obtain information about both charged and neutral particles. Three series of experiments were performed: (i) measurement of CL at different electron energies on samples of different thicknesses, (ii) measurements of TSL in combination with TSEE from pre-irradiated samples and (iii) recording of NsL curves at selected wavelengths during gradual heating of samples under a low current density electron beam.

**Results and discussion**

The CL spectrum of pure solid nitrogen recorded at 5 K in wide range from NIR up to VUV is shown in Fig. 1. Atomic transitions of nitrogen were observed in the entire range. In the VUV range we detected the 3s $^4$P → 2p$^3$ $^4$S transition at 120 nm. A distinctive feature of this line is its coincidence with the position in the gas phase spectrum within the accuracy of our measurements, which made it possible to attribute it to desorbing N(3s $^4$P) atoms [43]. The most intense in the spectrum was the so-called α-group near 523 nm, which corresponds to the $^2$D→$^4$S emission of N atoms initiated by matrix phonons. The less intense α'-group near 594.5 nm (not marked in the spectrum) represents the atomic $^2$D→$^4$S transition accompanied by simultaneous vibrational excitation of N$_2$ molecule ($v$=0 → $v$=1) in the electronic ground state [44]. The β-group stems from an oxygen which is invariably present as a minor impurity in the samples and corresponds to the $^1$S → $^3$P transition of O atom with satellites due to simultaneous vibration transitions of N$_2$ [24]. The γ-group under consideration is situated in the NIR range and consists of the γ-line at 794 nm and two red satellites distant from it by 125 and 250 cm$^{-1}$ [30]. The most intense molecular emissions covering the VUV and NUV ranges are the forbidden singlet molecular progression a'$^1\Sigma_u^-$ → X$^1\Sigma_g^+$ (system of Ogawa−Tanaka−Wilkinson−Mulliken) and the Vegard-Kaplan intercombination transition A$^3\Sigma_u^+$ → X$^1\Sigma_g^+$. The positions of the vibrational bands of both progressions are in good agreement with those measured earlier and reported in [18]. All the vibrational bands of both progressions show a matrix shift towards lower energy in comparison with the gas phase spectra indicating the bulk origin of the emitting species. The shift for the a'$^1\Sigma_u^-$ → X$^1\Sigma_g^+$ progression is 450 cm$^{-1}$ and 330 cm$^{-1}$ for the A$^3\Sigma_u^+$ → X$^1\Sigma_g^+$ one. The only molecular progression which shows no matrix shift is the second positive system – the transitions between C$^3\Pi_u$ and B$^3\Pi_g$ excited molecular states in the near UV range. As it was shown in [45] this emission belongs to N$_2$ molecules desorbing in the excited C$^3\Pi_u$ state.

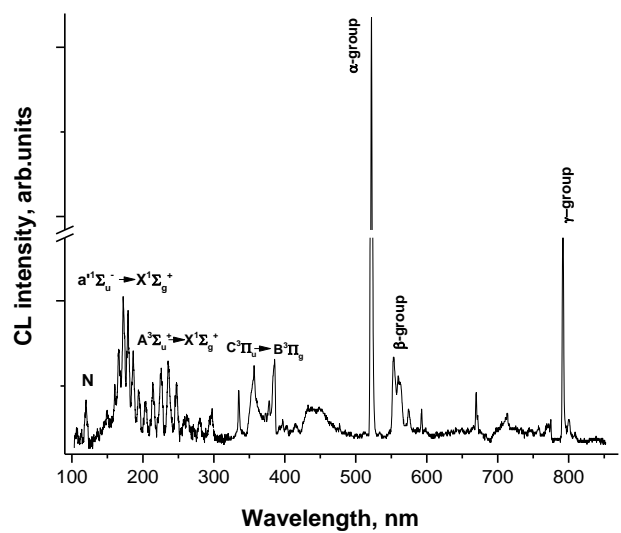

Fig. 1. Cathodoluminescence spectrum of solid nitrogen recorded at 5 K upon irradiation with a 0.5 keV electron beam.

In this spectrum we also detected the radiation-induced impurity center CN – the transition $B^2\Sigma^+ \to X^2\Sigma^+$, 0-0 line, near 388 nm. Some lines above 600 nm belong to the second spectral order.

We investigated the effect of excitation conditions on CL spectra with an emphasis on the visible and NIR ranges. Fig. 2 shows the CL spectra recorded at excitation energies $E_e$ of an electron beam $E_e = 0.5$ keV and 1.5 keV. Strong changes were observed in the range the second positive system. This system prevails at low electron beam energy and in thin samples, as it will be shown later, in contrast to the wide band background at 360 nm assigned to the $N_4$ ($D_{2h}$) structure of neutral $N_4$ [46]. It should be mentioned that this band was also observed in nitrogen nanoclusters immersed in liquid helium [47]. The α-group was more intense with respect to the γ-group at higher $E_e$: I(α)/I(γ)=4.1 at $E_e$=1.5 keV and I(α)/I(γ)=3.5 at $E_e$=0.5 keV. It is worthy of note that there are different channels of the excited N($^2$D) atom formation. It may appear as a result of direct excitation of N($^4$S) – product of $N_2$ molecule dissociation, under electron beam. Another possibility is the energy transfer from excited $N_2$ molecules in the $(A^3\Sigma_u^+)$ or $N_2(A^5\Sigma_g^+)$ state to N($^4$S) atom [48]. In Ref. [49], a new way of cold N($^2$D) atoms formation via the $2^5\Sigma_g^+$ state was suggested. One more channel of the excited N($^2$D) atom formation is the dissociative recombination of $N_3^+$ centers detected in an electron irradiated solid nitrogen ([18] and references therein):

$$N_3^+ + e^- \to N_2 + N + \Delta E$$

The energy release in this reaction comprises $\Delta E \sim 10.5$ eV [50] that provides possibility of N atom formation in the excited ($^2$D) state.

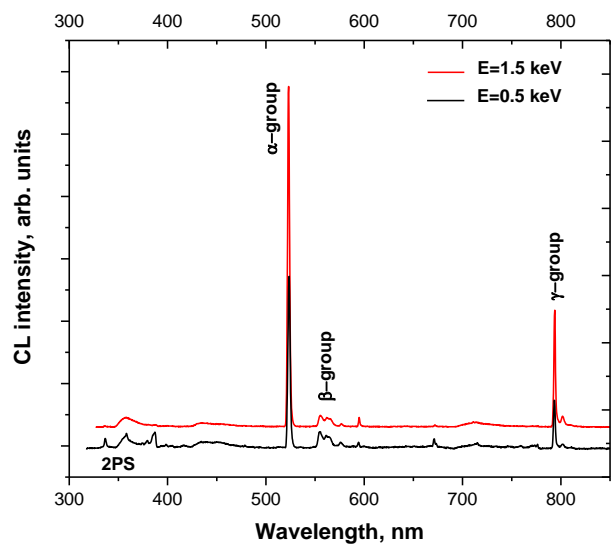

Fig. 2. CL spectra of solid nitrogen in the visible and NIR ranges recorded upon irradiation with 0.5 keV and 1.5 keV (red curve online) electron beam.

The change of nitrogen films thicknesses also resulted in change of the CL intensity spectral distribution. The CL spectra recorded for samples with a thickness of 30 nm and 200 μm are presented in Fig. 3. The characteristics of the second positive system are almost identical, but both the α-group and the γ-group are strongly suppressed in the thin film, indicating a bulk origin of both emitting centers. Their relative intensities are I(α)/I(γ)=3.2 for the film of 30 nm thickness and I(α)/I(γ)=3.5 for the 200 μm film. Note that the penetration depth d of electrons with an energy of 0.5 keV into solid nitrogen is 10 nm [51], so the thickness of both samples exceeds d and they accumulate an excessive negative charge [52], which creates favorable conditions for neutralization processes.

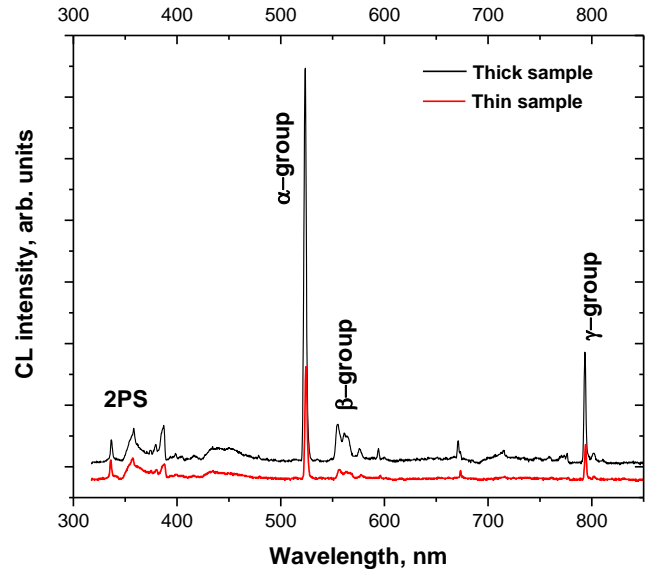

Fig. 3. CL spectra of solid nitrogen recorded for samples of 30 nm (red curve online) and 200 μm upon excitation with 0.5 keV electron beam at 5 K.

A comparison of the CL spectra obtained under different conditions as it is shown above indicates that there is no complete correlation in the behavior of the α and γ bands, which could be expected in the case of the formation of the γ-line emitting centers via electron attachment to the N($^2$D) atom.

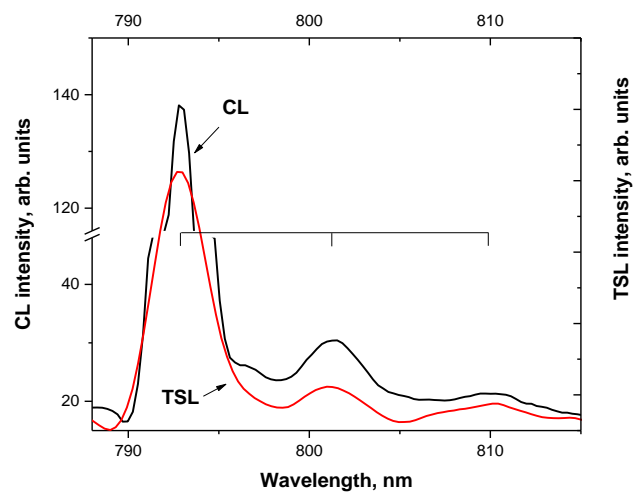

Fig. 4. Comparison of the γ-series in the CL spectrum recorded at 5 K with the TSL (red curve online) spectrum recorded at the TSL maximum at 16 K.

Increasing the irradiation time made it possible to register a new NIR band at 810 nm (the second satellite of the γ-line) in the spectra of TSL. The positions of all three spectral features coincide in the spectra of spontaneous and stimulated luminescence, as shown in Fig. 4.

The glow curves were measured for all three components of the γ-series. As can be seen from Fig. 5, they correlate with each other indicating a common origin of these bands.

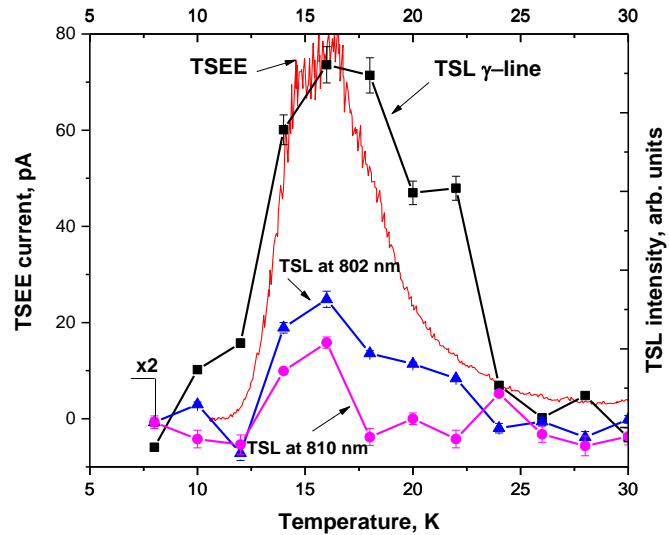

Fig. 5. TSL at the wavelengths of the components of the γ-group structure and TSEE yield.

Moreover, they all correlate with the thermally stimulated exoelectron emission TSEE, that points to their connection with the neutralization reaction stimulated by electron detrapping. Due to the negative electron affinity of solid nitrogen (-1.8 eV) [53], which results in the absence of a barrier to electron escape, no shift was observed between the TSL and TSEE yield maxima. As it was found in [30] the TSL detected at the γ-line correlates with the TSL recorded at the β-group near 560 nm connected with ($^1S \to ^3P$) transition of O atom characterized by a very high positive electron affinity χ = 1.46 eV [54], i.e even in this case the neutralization channel of relaxation appeared to be predominant. The primary intrinsic positive charge centers in solid nitrogen are the self-trapped hole. This is indicated by low mobility of positive charge carriers in solid nitrogen [55]. The dimerization reaction $N_2^+ + N_2 \to N_4^+$ with $N_4^+$ formation was detected in supersonic nitrogen jet below 20 K [56]. Experiments on electronically stimulated desorption from solid nitrogen [57, 58] provided additional evidence for the formation of $N_4^+$ via the dimerization reaction. In our previous study [35] it was shown that tetranitrogen cation $N_4^+$ manifests itself by the dissociative recombination reaction with electron (reaction 5) in the TSL and NsL experiments. In these experiments stimulated emission was detected at wavelengths corresponding to transitions from several vibronic levels of the $a'^1\Sigma_u^-$ state. An appearance of the excited $N_2$ molecules upon recombination of $N_4^+$ with electron was explained by an absence of the barrier for the state with dissociation limit $N_2(a'^1\Sigma_u^-) + N_2(a'^1\Sigma_u^-)$ according the theoretical study of the ground and excited states of $N_4(D_{2h})$ [59]. We carried out measurements of nonstationary luminescence at a wavelength of the 0−4 band of the a'$^1\Sigma_u^-$ → X$^1\Sigma_g^+$ molecular transition and the NsL at the γ-line (Fig.6).

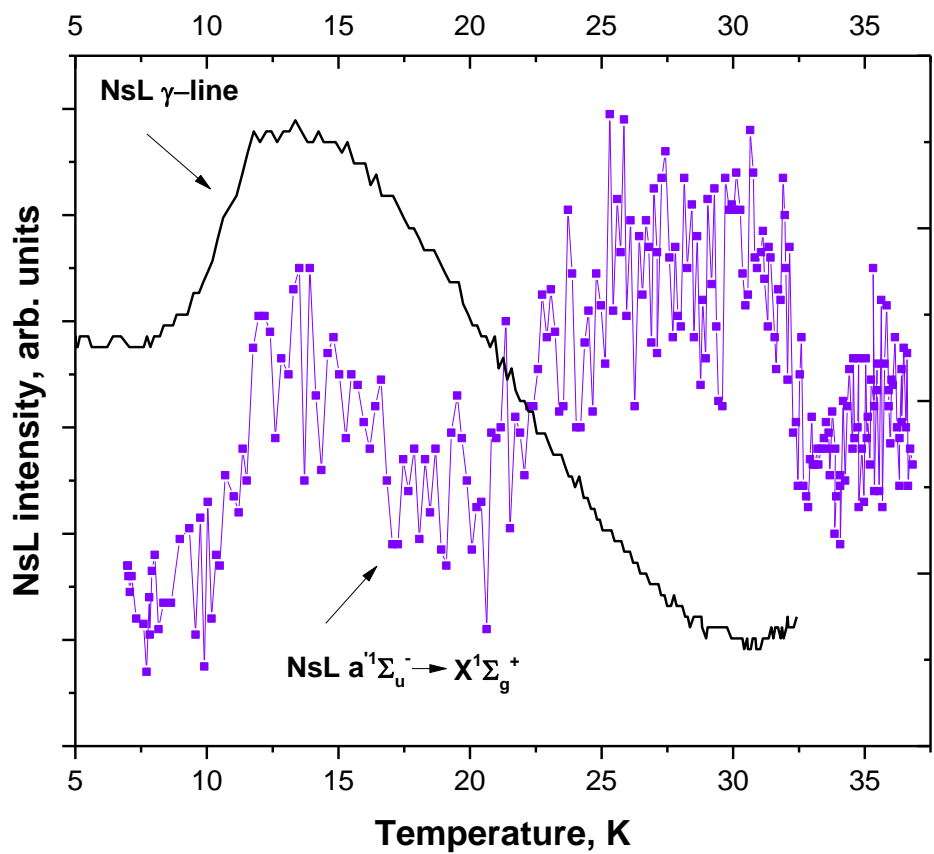

Fig. 6. Nonstationary luminescence curves measured at the γ-line and at the 0-4 band of the a'$^1\Sigma_u^-$ → X$^1\Sigma_g^+$ transition of N$_2$ molecule.

Both curves correlate in the low temperature range from 7 up to 20 K and their maxima coincide. Note, that in this experiment we used lower speed of heating – 3 Kmin$^{-1}$, that resulted in a shift of the NsL maxima to lower temperature with respect to the TSL and TSEE maxima shown in Fig. 5. The NsL curve for the γ–line has no features in the range of temperature above 20 K, it decreases monotonically following the TSEE yield. The correlation of the γ–line NsL in the range of low temperatures with the NsL measured at the 0-4 band of the a'$^1\Sigma_u^-$ → X$^1\Sigma_g^+$ transition, which is the "marker" of the tetranitrogen cation N$_4^+$ dissociative recombination [35], indicates possible connection of the γ–line with the neutralization of N$_4^+$, which proceeds however *via* the „cage effect" scenario – reaction (6). The probability of scenarios is determined by the presence of barriers to dissociation. As it was found in [38] the lowest optically accessible excited state $1^1B_{3u}$ of isomer N$_4$ (D$_{2h}$) lies 1.57 eV above the ground state $^1A_g$. A barrier to dissociation of this state is about 6.5 kcal/mol, that prevents dissociation of isomer in this state at low temperatures and creates favorable conditions for the „cage effect" scenario of the neutralization reaction N$_4^+$ + e$^-$. The estimated energy of the radiative transition $1^1B_{3u}$ → $^1A_g$ of isomer N$_4$ of D$_{2h}$ configuration is about 1.55 eV, that is very close to the experimentally measured position of the γ–line in both spontaneous and stimulated luminescence (1.563 eV). However, the experimental harmonic frequency (125 cm$^{-1}$) of the γ-series in spontaneous/stimulated luminescence is much less than the lowest calculated one (469 cm$^{-1}$) for the N$_4$ (D$_{2h}$) isomer [38]. The experimental frequency is close to that for open-chain structures N$_4$ $C_s$ [13, 60] and coincides with the calculated frequency of ionic form N$_4^+$ C$_s$ (123 cm$^{-1}$ according [60]). The structure of N$_4^+$ in the Ne matrix was determined by ESR [61] and IR [62] studies. It was established that in this case the N$_4^+$ structure corresponds to the open chain configuration D$_{\infty h}$. Recently, N$_4^+$ was found in the pure N$_2$ solid irradiated with VUV photons [63]. Its vibrational frequency turned out to be almost identical to the gas-phase and Ne-matrix values. It suggests that the N$_4^+$ in solid nitrogen has the same structure D$_{\infty h}$ as in the Ne matrix. However, the structure registered in our experiments with solid nitrogen is not a van der Waals complex, the measured frequency is 5 times higher than that characteristic of van der Waals structures [60] (supporting information).

Interestingly that the behavior of NsL curves for the γ–line and molecular a'-X transition changes to the opposite in the range of temperatures higher than 20 K where thermally activated diffusion processes come into play. Nitrogen atom in the ground state N($^4$S) can be located in three trapping sites in nitrogen matrix [64, 65]: (s) substitutional site, (i$_{oh}$) interstitial octahedral distorted site and (i$_{th}$) interstitial tetrahedral distorted site. As a rule, atoms in substitutional sites are most stable while atoms located in interstitial sites have lower activation energy for thermally driven diffusion [66]. The second peak in NsL of the a'-X transition at about 28 K has no analogues either in the NsL γ-line or in the TSEE yield. The question arises as to how the a'$^1\Sigma_u^-$ state observed in NsL is populated at higher temperatures (T > 20 K). The a'$^1\Sigma_u^-$ state is a lowest singlet excited state, and its dissociation limit is formed by two nitrogen atoms in excited states: N($^2$D) + N($^2$D) [67]. As mentioned above, excited N($^2$D) atoms can be formed in a number of processes: (i) by direct excitation of N($^4$S) with an electron beam; (ii) by energy transfer from the N$_2$ (A$^3\Sigma_u^+$) or N$_2$(A$^5\Sigma_g^+$) state to the N($^4$S) atom [48]; (iii) via the 2$^5\Sigma_g^+$ state [49]; (iv) via the dissociative recombination of N$_3^+$ [18]. During NsL measurements, the contribution from channel (i) is very low because these measurements are performed at low current densities. Channel (ii) manifests itself in the low level of A-X emission in the TSL spectra of solid nitrogen together with the high intensity of the α-group [22]. The contribution of channel (iii) has not been studied so far. Under the conditions of NsL and TSL measurements, the channel (iv) of $^2$D atom formation via dissociative recombination does not operate in the temperature range of 25-30 K since there are no electron traps with corresponding activation energies, as follows from the TSEE analysis. The contribution of close pair recombination at higher temperatures seems unlikely. Considering the role of these channels in the "high temperature" NsL at the a'-X transition, we can conclude that the initial process of $^2$D atom formation is most likely the thermally activated diffusion of $^4$S atoms followed by their recombination and energy transfer by channel (ii). The absence of structure on the NsL curve for a'-X emission in the temperature range of 20–30 K is due to electron traps at structural defects in addition to s, i$_{oh}$, and i$_{th}$ trapping sites in the ideal lattice, since the measurements were carried out on an unannealed sample. The peak about 35 K is due to the α-β phase transition of solid nitrogen, which was observed in an early TSL study by Brian Brocklehurst and George C. Pimentel [68] and later studied by the TSEE technique [40]. Thus, the NsL method enabled us to monitor both the charge recombination and the recombination of neutrals. Herewith, a'-X emission occurs as a result of two different processes – the reaction of neutralization of $N_4^+$ at low temperatures and atomic diffusion with subsequent formation of $^2D$ atoms and their recombination at temperatures above 20 K. The completely different behavior of the γ-line NsL and the "marker" of $^2D$ atom formation (the a'-X emission) in the temperature range above 20 K excludes the attribution of the γ-line to $N^-$ anion emission. On the other hand, the correlation of the γ–line NsL in the range of low temperatures (5-20 K) with the NsL measured at the 0-4 band of the $a'^1\Sigma_u^- \rightarrow X^1\Sigma_g^+$ transition, which is connected with the cation $N_4^+$ neutralization in irradiated solid nitrogen [35], supports the possible connection of the γ–line with the neutralization of $N_4^+$, proceeding according the „cage effect" scenario with the formation of tetranitrogen. We hope that the results presented in this article will encourage further experimental and theoretical research in this direction.

**Summary**

This work is a development of our studies of relaxation processes in solid nitrogen irradiated with electrons of subthreshold energy [30] with a focus on the γ–line origin. The relaxation dynamics was monitored by emission spectroscopy – cathodoluminescence (CL) and nonstationary luminescence (NsL), along with optical and current activation spectroscopy: TSL and TSEE. A new satellite of the γ-line in stimulated luminescence has been discovered at 810 nm. The positions of all three spectral features which form the γ–group coincide in the spectra of spontaneous and stimulated luminescence, as evidenced by a comparison of the CL spectrum recorded at 5 K with the TSL spectrum recorded at the TSL maximum at 16 K. The similar behavior of all these lines in the stimulated luminescence (TSL) and their correlation with the stimulated current (TSEE) indicate common origin of these lines and their connection with the neutralization reaction. The detection of a second satellite of the γ-line in the spectra of spontaneous and stimulated luminescence casts doubt on the identification of the γ-line as emission from the nitrogen anion $N^-$ [29] because satellites did not figure in their model. A comparison of the CL spectra obtained under different conditions showed that there is no correlation in the behavior of emission from the $^2D$ state of the N atom and the γ-line, which could be expected in the case of the formation of the γ-line emitting centers via electron attachment to the $N(^2D)$ atom. The found correlation of the γ–line NsL with the NsL of the $N_4^+$ dissociative recombination product $N_2 (a'^1\Sigma_u^-)$ indicates a probable connection of the γ–series with $N_4$. The measured $E$=1.56 eV of the γ–line in spontaneous and stimulated luminescence is close to the predicted one $E$=1.55 eV for $N_4$ of $D_{2h}$ configuration [38], but the difference in harmonic frequencies still leaves the question of the origin of the γ-group, which requires further experimental and theoretical research.

**Acknowledgements**

The authors cordially thank Tore Brinck, Minh Nguyen, Yu-Jong Wu, Alec Wodtke, Paul Mayer, Roman Boltnev and Vladimir Khmelenko for sharing relevant data and discussions.